\documentclass[aip,graphicx]{revtex4-1}
\usepackage [latin1]{inputenc}
\draft 

\begin{document}

\title{The equivalence between local inertial frames and electromagnetic gauge in Einstein-Maxwell theories} 

\author{Alcides Garat}
\affiliation{Former Professor at Universidad de la Rep\'{u}blica, Av. 18 de Julio 1824-1850, 11200 Montevideo, Uruguay.}
\email[]{garat.alcides@gmail.com}
\date{\today}

\begin{abstract}
\begin{center}
{\bf Abstract}
\end{center}
It has been proven that locally the inertial frames and gauge states of the electromagnetic field are equivalent. This proof is valid for Einstein-Maxwell theories in four-dimensional Lorentzian spacetimes. Use will be made of theorems proved in a previous manuscript. These theorems state that locally the group of electromagnetic gauge transformations is isomorphic to the local group of Lorentz transformations of a special set of tetrad vectors. The tetrad that locally and covariantly diagonalizes any non-null electromagnetic stress-energy tensor. Two isomorphisms, one for each orthogonal plane of stress-energy eigenvectors. We will discuss the opposite problem in this paper. What happens with local electromagnetic gauge when the test object under study is boosted by any mechanical means. We will prove that boosting matter is indistinguishable from introducing an appropriate local electromagnetic gauge transformation.
\end{abstract}



\maketitle 

\section{Introduction}
\label{intro}

In manuscript \cite{A} a covariant method for the local diagonalization of the $U(1)$ electromagnetic stress-energy tensor was presented. At every point in a curved four dimensional Lorentzian spacetime a new tetrad was introduced for non-null electromagnetic fields such that this tetrad locally and covariantly diagonalizes the stress-energy tensor. At every point the timelike and one spacelike vectors generate a plane that we called blade one \cite{A,ROMP,SING,ATGU,GLAW,IWCP,LomCon,AEO,AMONO,SCR,FULL,SCH}. The other two spacelike vectors generate a plane that we called blade two. These vectors are constructed with the local extremal field \cite{MW}, its dual, the very metric tensor and a pair of vector fields that represent a generic choice as long as the tetrad vectors do not become trivial. Let us display for the Abelian case the explicit expression for these vectors,

\begin{eqnarray}
U^{\alpha} &=& \xi^{\alpha\lambda}\:\xi_{\rho\lambda}\:X^{\rho} \:
/ \: (\: \sqrt{-Q/2} \: \sqrt{X_{\mu} \ \xi^{\mu\sigma} \
\xi_{\nu\sigma} \ X^{\nu}}\:) \label{U}\\
V^{\alpha} &=& \xi^{\alpha\lambda}\:X_{\lambda} \:
/ \: (\:\sqrt{X_{\mu} \ \xi^{\mu\sigma} \
\xi_{\nu\sigma} \ X^{\nu}}\:) \label{V}\\
Z^{\alpha} &=& \ast \xi^{\alpha\lambda} \: Y_{\lambda} \:
/ \: (\:\sqrt{Y_{\mu}  \ast \xi^{\mu\sigma}
\ast \xi_{\nu\sigma}  Y^{\nu}}\:)
\label{Z}\\
W^{\alpha} &=& \ast \xi^{\alpha\lambda}\: \ast \xi_{\rho\lambda}
\:Y^{\rho} \: / \: (\:\sqrt{-Q/2} \: \sqrt{Y_{\mu}
\ast \xi^{\mu\sigma} \ast \xi_{\nu\sigma} Y^{\nu}}\:) \ .
\label{W}
\end{eqnarray}

We will explain in detail how to obtain the tetrad (\ref{U}-\ref{W}) and will also study its properties. Throughout the paper we use the conventions of manuscript \cite{MW}. In particular we use a metric with sign conventions -+++. The only difference in notation with \cite{MW} will be that we will call our geometrized electromagnetic potential $A^{\alpha}$, where $f_{\mu\nu}=A_{\nu ;\mu} - A_{\mu ;\nu}$ is the geometrized electromagnetic field $f_{\mu\nu}= (G^{1/2} / c^2) \: F_{\mu\nu}$. We start by stating that at every point in spacetime there is a duality rotation by an angle $-\alpha$ that transforms a non-null electromagnetic field into an extremal field,

\begin{equation}
\xi_{\mu\nu} = e^{-\ast \alpha} f_{\mu\nu}\ = \cos(\alpha)\:f_{\mu\nu} - \sin(\alpha)\:\ast f_{\mu\nu}.\label{dref}
\end{equation}

where  $\ast f_{\mu\nu}={1 \over 2}\:\epsilon_{\mu\nu\sigma\tau}\:f^{\sigma\tau}$ is the dual tensor of $f_{\mu\nu}$ and $\epsilon_{\mu\nu\sigma\tau}$ is the alternating tensor. We recall that the Levi-Civita pseudotensor can be transformed into a tensor through the use of factors $\sqrt{-g}$, where $g$ is the determinant of the metric tensor, see reference \cite{A}. The local scalar $\alpha$ is known as the complexion of the electromagnetic field. It is a local gauge invariant quantity. Extremal fields are essentially electric fields and they satisfy,

\begin{equation}
\xi_{\mu\nu} \ast \xi^{\mu\nu}= 0\ . \label{i0}
\end{equation}

Equation (\ref{i0}) is a condition imposed on (\ref{dref}) and then the explicit expression for the complexion is found to be $\tan(2\alpha) = - f_{\mu\nu}\:\ast f^{\mu\nu} / f_{\lambda\rho}\:f^{\lambda\rho}$. As antisymmetric fields in a four dimensional Lorentzian spacetime, the extremal fields also verify the identity,

\begin{eqnarray}
\xi_{\mu\alpha}\:\xi^{\nu\alpha} - \ast \xi_{\mu\alpha}\: \ast \xi^{\nu\alpha} &=& \frac{1}{2} \: \delta_{\mu}^{\:\:\:\nu}\ Q
\ ,\label{i1}
\end{eqnarray}

where $Q=\xi_{\mu\nu}\:\xi^{\mu\nu}=-\sqrt{T_{\mu\nu}T^{\mu\nu}}$ according to equations (39) in \cite{MW}. $Q$ is assumed not to be zero, because we are dealing with non-null electromagnetic fields.  Non-null we clarify means basically that $f_{\mu\nu}\:f^{\mu\nu}\neq0$ and $\ast f_{\mu\nu}\:f^{\mu\nu}\neq0$. In turn and by definitions these last equations imply that $\xi_{\mu\nu}\:\xi^{\mu\nu}\neq0$. It can be proved that condition (\ref{i0}) together with the general identity,

\begin{eqnarray}
A_{\mu\alpha}\:B^{\nu\alpha} -
\ast B_{\mu\alpha}\: \ast A^{\nu\alpha} &=& \frac{1}{2}
\: \delta_{\mu}^{\:\:\:\nu}\: A_{\alpha\beta}\:B^{\alpha\beta}  \ ,\label{ig}
\end{eqnarray}

which is valid for every pair of antisymmetric tensors in a four-dimensional Lorentzian spacetime \cite{MW}, when applied to the case $A_{\mu\alpha} = \xi_{\mu\alpha}$ and $B^{\nu\alpha} = \ast \xi^{\nu\alpha}$ yields the equivalent condition,

\begin{eqnarray}
\xi_{\alpha\mu}\:\ast \xi^{\mu\nu} &=& 0\ ,\label{i2}
\end{eqnarray}

which is equation (64) in \cite{MW}. The duality rotation given by equation (59) in\cite{MW} which is the inverse of equation (\ref{dref}),

\begin{equation}
f_{\mu\nu} = \xi_{\mu\nu} \: \cos\alpha + \ast\xi_{\mu\nu} \: \sin\alpha\ ,\label{dr}
\end{equation}

allows us to express the stress-energy tensor in terms of the extremal field,

\begin{equation}
T_{\mu\nu}=\xi_{\mu\lambda}\:\:\xi_{\nu}^{\:\:\:\lambda}
+ \ast \xi_{\mu\lambda}\:\ast \xi_{\nu}^{\:\:\:\lambda}\ .\label{TEMDR}
\end{equation}

There are four tetrad vectors that at every point in spacetime diagonalize the stress-energy tensor (\ref{TEMDR}) in geometrodynamics,

\begin{eqnarray}
V_{(1)}^{\alpha} &=& \xi^{\alpha\lambda}\:\xi_{\rho\lambda}\:X^{\rho}
\label{V1}\\
V_{(2)}^{\alpha} &=& \sqrt{-Q/2} \: \xi^{\alpha\lambda} \: X_{\lambda}
\label{V2}\\
V_{(3)}^{\alpha} &=& \sqrt{-Q/2} \: \ast \xi^{\alpha\lambda} \: Y_{\lambda}
\label{V3}\\
V_{(4)}^{\alpha} &=& \ast \xi^{\alpha\lambda}\: \ast \xi_{\rho\lambda}
\:Y^{\rho}\ .\label{V4}
\end{eqnarray}

When we make iterative use of (\ref{i1}) and (\ref{i2}) we find,

\begin{eqnarray}
V_{(1)}^{\alpha}\:T_{\alpha}^{\:\:\:\beta} &=& \frac{Q}{2}\:V_{(1)}^{\beta}
\label{EV1}\\
V_{(2)}^{\alpha}\:T_{\alpha}^{\:\:\:\beta} &=& \frac{Q}{2}\:V_{(2)}^{\beta}
\label{EV2}\\
V_{(3)}^{\alpha}\:T_{\alpha}^{\:\:\:\beta} &=& -\frac{Q}{2}\:V_{(3)}^{\beta}
\label{EV3}\\
V_{(4)}^{\alpha}\:T_{\alpha}^{\:\:\:\beta} &=& -\frac{Q}{2}\:V_{(4)}^{\beta}\ .
\label{EV4}
\end{eqnarray}

With all these elements it becomes trivial to prove that the tetrad \cite{WE,MTW} (\ref{V1}-\ref{V4}) is orthogonal and diagonalizes the stress-energy tensor (\ref{TEMDR}). We notice then that we still have to define the vectors $X^{\mu}$ and $Y^{\mu}$. Let us introduce some nomenclature. The tetrad vectors have two fundamental structural components. For instance in vector $U^{\alpha}$ equation (\ref{U}) there are two main structures. First, the skeleton, in this case $\xi^{\alpha\lambda}\:\xi_{\rho\lambda}$, and second, the gauge vector $X^{\rho}$. These do not include the normalization factor $1 / \: (\: \sqrt{-Q/2} \: \sqrt{X_{\mu} \ \xi^{\mu\sigma} \ \xi_{\nu\sigma} \ X^{\nu}}\:)$. The gauge vectors it was proved in manuscript \cite{A} could be anything that will not make the tetrad vectors trivial. That is, the tetrad (\ref{U}-\ref{W}) diagonalizes the stress-energy tensor for any non-trivial gauge vectors $X^{\mu}$ and $Y^{\mu}$. It was therefore proved that we can make different choices for $X^{\mu}$ and $Y^{\mu}$. In geometrodynamics, the Maxwell equations,

\begin{eqnarray}
f^{\mu\nu}_{\:\:\:\:\:;\nu} &=& 0 \label{L1}\nonumber\\
\ast f^{\mu\nu}_{\:\:\:\:\:;\nu} &=& 0 \ , \label{L2}
\end{eqnarray}

reveal that two potential vector fields $A_{\nu}$ and $\ast A_{\nu}$ exist \cite{CF},

\begin{eqnarray}
f_{\mu\nu} &=& A_{\nu ;\mu} - A_{\mu ;\nu}\label{ER}\nonumber\\
\ast f_{\mu\nu} &=& \ast A_{\nu ;\mu} - \ast A_{\mu ;\nu} \ .\label{DER}
\end{eqnarray}

The symbol $``;''$ stands for covariant derivative with respect to the metric tensor $g_{\mu\nu}$ and the star in $\ast A_{\nu}$ is just a name, not the dual Hodge operator, meaning that $\ast A_{\nu ;\mu} = (\ast A_{\nu})_{;\mu}$. We can define then, a tetrad,

\begin{eqnarray}
U^{\alpha} &=& \xi^{\alpha\lambda}\:\xi_{\rho\lambda}\:A^{\rho} \:
/ \: (\: \sqrt{-Q/2} \: \sqrt{A_{\mu} \ \xi^{\mu\sigma} \
\xi_{\nu\sigma} \ A^{\nu}}\:) \label{UO}\\
V^{\alpha} &=& \xi^{\alpha\lambda}\:A_{\lambda} \:
/ \: (\:\sqrt{A_{\mu} \ \xi^{\mu\sigma} \
\xi_{\nu\sigma} \ A^{\nu}}\:) \label{VO}\\
Z^{\alpha} &=& \ast \xi^{\alpha\lambda} \: \ast A_{\lambda} \:
/ \: (\:\sqrt{\ast A_{\mu}  \ast \xi^{\mu\sigma}
\ast \xi_{\nu\sigma}  \ast A^{\nu}}\:)
\label{ZO}\\
W^{\alpha} &=& \ast \xi^{\alpha\lambda}\: \ast \xi_{\rho\lambda}
\:\ast A^{\rho} \: / \: (\:\sqrt{-Q/2} \: \sqrt{\ast A_{\mu}
\ast \xi^{\mu\sigma} \ast \xi_{\nu\sigma} \ast A^{\nu}}\:) \ .
\label{WO}
\end{eqnarray}

The four vectors (\ref{UO}-\ref{WO}) have the following algebraic properties,

\begin{equation}
-U^{\alpha}\:U_{\alpha}=V^{\alpha}\:V_{\alpha}
=Z^{\alpha}\:Z_{\alpha}=W^{\alpha}\:W_{\alpha}=1 \ .\label{TSPAUX}
\end{equation}

Using the equations (\ref{i1}-\ref{i2}) it is simple to prove that (\ref{UO}-\ref{WO}) are orthogonal and normalized. When we make the transformation,

\begin{eqnarray}
A_{\alpha} \rightarrow A_{\alpha} + \Lambda_{,\alpha}\ , \label{G1}
\end{eqnarray}

$f_{\mu\nu}$ remains invariant, and the transformation,

\begin{eqnarray}
\ast A_{\alpha} \rightarrow \ast A_{\alpha} +
\ast \Lambda_{,\alpha}\ , \label{G2}
\end{eqnarray}

leaves $\ast f_{\mu\nu}$ invariant, as long as the functions $\Lambda$ and $\ast \Lambda$ are local scalars. We emphasize that $\ast \Lambda$ is just a local scalar function that might be different from $\Lambda$, that is, the star in $\ast \Lambda$ is just a name, not the Hodge operator.  Schouten defined what he called, a two-bladed structure in a spacetime \cite{SCH}. These blades are the planes determined by the pairs
($U^{\alpha}, V^{\alpha}$) and ($Z^{\alpha}, W^{\alpha}$). It was proved in \cite{A,ROMP,SING} that the transformation (\ref{G1}) generates a ``hyperbolic rotation'' of the tetrad vectors ($U^{\alpha}, V^{\alpha}$) into ($\tilde{U}^{\alpha}, \tilde{V}^{\alpha}$) such that these ``rotated'' vectors ($\tilde{U}^{\alpha}, \tilde{V}^{\alpha}$) remain in the plane or blade one generated or spanned by ($U^{\alpha}, V^{\alpha}$). It was also proved in \cite{A} that the transformation (\ref{G2}) generates a ``spatial rotation'' of the tetrad vectors ($Z^{\alpha}, W^{\alpha}$) into ($\tilde{Z}^{\alpha}, \tilde{W}^{\alpha}$) such that these ``rotated'' vectors ($\tilde{Z}^{\alpha}, \tilde{W}^{\alpha}$) remain in the plane or blade two generated by ($Z^{\alpha}, W^{\alpha}$).  For the sake of simplicity we will assume that the transformation of the two vectors $(U^{\alpha},\:V^{\alpha})$ on blade one, given in (\ref{UO}-\ref{VO}), by the ``angle'' $\phi$ is a proper transformation, that is, a boost. For discrete proper and improper transformations the result follows the same lines \cite{A,ROMP,SING}. Therefore we can write the transformation generated by (\ref{G1}) as,

\begin{eqnarray}
U^{\alpha}_{(\phi)}  &=& \cosh(\phi)\: U^{\alpha} +  \sinh(\phi)\: V^{\alpha} \label{UT} \\
V^{\alpha}_{(\phi)} &=& \sinh(\phi)\: U^{\alpha} +  \cosh(\phi)\: V^{\alpha} \label{VT} \ .
\end{eqnarray}

The transformation generated by (\ref{G2}) of the two tetrad vectors $(Z^{\alpha},\:W^{\alpha})$ on blade two, given in (\ref{ZO}-\ref{WO}), by the ``angle'' $\varphi$, can be expressed as,

\begin{eqnarray}
Z^{\alpha}_{(\varphi)}  &=& \cos(\varphi)\: Z^{\alpha} -  \sin(\varphi)\: W^{\alpha} \label{ZT} \\
W^{\alpha}_{(\varphi)}  &=& \sin(\varphi)\: Z^{\alpha} +  \cos(\varphi)\: W^{\alpha} \label{WT} \ .
\end{eqnarray}

It is a simple exercise in algebra to see that the equalities $U^{[\alpha}_{(\phi)}\:V^{\beta]}_{(\phi)} = U^{[\alpha}\:V^{\beta]}$ and $Z^{[\alpha}_{(\varphi)}\:W^{\beta]}_{(\varphi)} = Z^{[\alpha}\:W^{\beta]}$ are true. These equalities are telling us that these antisymmetric tetrad objects are gauge invariant. We remind ourselves that it was proved in manuscripts \cite{A,ROMP} that the group of local electromagnetic gauge transformations is isomorphic to the group LB1 of boosts plus discrete transformations on blade one, and independently to LB2, the group of spatial rotations on blade two. LB1 is the group composed by the boosts in the local plane one $SO(1,1)$ and two discrete transformations. One of the discrete transformations is minus the identity two by two, also known as the full inversion. The other discrete transformation is given by $\Lambda^{o}_{\:\:o} = 0$, $\Lambda^{o}_{\:\:1} = 1$, $\Lambda^{1}_{\:\:o} = 1$, $\Lambda^{1}_{\:\:1} = 0$ also known as switch or flip. We notice that this discrete transformation is not a Lorentz transformation because it is a reflection. The group LB2 is just $SO(2)$ in the local orthogonal plane two. Equations (\ref{UT}-\ref{VT}) represent a local electromagnetic gauge transformation of the vectors $(U^{\alpha}, V^{\alpha})$. Equations (\ref{ZT}-\ref{WT}) represent a local electromagnetic gauge transformation of the vectors $(Z^{\alpha}, W^{\alpha})$.

The two gauge vectors $X^{\alpha}$ and $Y^{\alpha}$ are spacetime gauge, spacetime freedom. Gauge is not only a property of electromagnetic fields, it is clearly through the above, a property of spacetime itself. Once we made the choice $X^{\alpha}=A^{\alpha}$ and $Y^{\alpha}=\ast A^{\alpha}$ in the normalized tetrad (\ref{UO}-\ref{WO}) the question about the geometrical implications of electromagnetic gauge transformations of the tetrad vectors (\ref{UO}-\ref{WO}) arises. We first notice that a local electromagnetic gauge transformation of the ``gauge vectors'' $X^{\alpha}=A^{\alpha}$ and $Y^{\alpha}=\ast A^{\alpha}$ can be just interpreted as a new choice for the gauge vectors $X_{\alpha} = A_{\alpha} + \Lambda_{,\alpha}$ and $Y_{\alpha} = \ast A_{\alpha} + \ast \Lambda_{,\alpha}$ in tetrad (\ref{V1}-\ref{V4}) or its normalized version. When we make the transformation, $A_{\alpha} \rightarrow A_{\alpha} + \Lambda_{,\alpha}$, $f_{\mu\nu}$ remains invariant, and the transformation, $\ast A_{\alpha} \rightarrow \ast A_{\alpha} + \ast \Lambda_{,\alpha}$, leaves $\ast f_{\mu\nu}$ invariant, as long as the functions $\Lambda$ and $\ast \Lambda$ are local scalars. It is valid to ask how the tetrad vectors (\ref{V1}-\ref{V2}) will transform under $A_{\alpha} \rightarrow A_{\alpha} + \Lambda_{,\alpha}$ and (\ref{V3}-\ref{V4}) under $\ast A_{\alpha} \rightarrow \ast A_{\alpha} + \ast \Lambda_{,\alpha}$. Schouten defined what he called, a two-bladed structure in a spacetime \cite{SCH}. These local blades or planes are the planes determined by the pairs ($V_{(1)}^{\alpha}, V_{(2)}^{\alpha}$) in the case of the local plane or blade one and ($V_{(3)}^{\alpha}, V_{(4)}^{\alpha}$) in the case of the local orthogonal plane or blade two. Given the brief nature of this introduction we will limit ourselves to show a few illustrative results as far as tetrad transformations for gauge vector choice given by electromagnetic gauge transformations. The whole analysis is given in manuscripts \cite{A,ROMP}. In order to simplify the notation we will write $\Lambda_{,\alpha}=\Lambda_{\alpha}$. First we study the change in (\ref{V1}-\ref{V2}) under $A_{\alpha} \rightarrow A_{\alpha} + \Lambda_{,\alpha}$. Using the following notation, $C=(-Q/2)\:V_{(1)\sigma}\:\Lambda^{\sigma} / (\:V_{(2)\beta}\:V_{(2)}^{\beta}\:)$ and $D=(-Q/2)\:V_{(2)\sigma}\:\Lambda^{\sigma} /(\:V_{(1)\beta}\:V_{(1)}^{\beta}\:)$, several cases arise on blade one. $C$ and $D$ are local scalars that have been chosen in their present form in order to make the local Lorentz transformations in equations (\ref{TN1}-\ref{TN2}) more evident, specially with regards to local group transformations as in reference \cite{GLAW}. We would like to calculate the norm of the transformed vectors $\tilde{V}_{(1)}^{\alpha}$ and $\tilde{V}_{(2)}^{\alpha}$, see the detailed analysis in reference \cite{A},

\begin{eqnarray}
\tilde{V}_{(1)}^{\alpha}\:\tilde{V}_{(1)\alpha} &=&
[(1+C)^2-D^2]\:V_{(1)}^{\alpha}\:V_{(1)\alpha}\label{FP}\\
\tilde{V}_{(2)}^{\alpha}\:\tilde{V}_{(2)\alpha} &=&
[(1+C)^2-D^2]\:V_{(2)}^{\alpha}\:V_{(2)\alpha}\ ,\label{SP}
\end{eqnarray}

where the relation $V_{(1)}^{\alpha}\:V_{(1)\alpha}= -V_{(2)}^{\alpha}\:V_{(2)\alpha}$ has been used and $V_{(1)}^{\alpha}$ assumed timelike for simplicity. In order for these transformations to keep the timelike or spacelike character of $V_{(1)}^{\alpha}$ and $V_{(2)}^{\alpha}$ the condition $[(1+C)^2-D^2]>0$ must be satisfied. If this condition is
fulfilled, then we can normalize the transformed vectors $\tilde{V}_{(1)}^{\alpha}$ and $\tilde{V}_{(2)}^{\alpha}$ as follows,

\begin{eqnarray}
{\tilde{V}_{(1)}^{\alpha}
\over \sqrt{-\tilde{V}_{(1)}^{\beta}\:\tilde{V}_{(1)\beta}}}&=&
{(1+C) \over \sqrt{(1+C)^2-D^2}}
\:{V_{(1)}^{\alpha} \over \sqrt{-V_{(1)}^{\beta}\:V_{(1)\beta}}}+
{D \over \sqrt{(1+C)^2-D^2}}
\:{V_{(2)}^{\alpha} \over \sqrt{V_{(2)}^{\beta}\:V_{(2)\beta}}}\label{TN1}\\
{\tilde{V}_{(2)}^{\alpha}
\over \sqrt{\tilde{V}_{(2)}^{\beta}\:\tilde{V}_{(2)\beta}}}&=&
{D \over \sqrt{(1+C)^2-D^2}}
\:{V_{(1)}^{\alpha} \over \sqrt{-V_{(1)}^{\beta}\:V_{(1)\beta}}} +
{(1+C) \over \sqrt{(1+C)^2-D^2}}
\:{V_{(2)}^{\alpha} \over \sqrt{V_{(2)}^{\beta}\:V_{(2)\beta}}}\ .
\label{TN2}
\end{eqnarray}

$U^{\alpha} = {V_{(1)}^{\alpha} \over \sqrt{-V_{(1)}^{\beta}\:V_{(1)\beta}}}$ and $V^{\alpha} = {V_{(2)}^{\alpha} \over \sqrt{V_{(2)}^{\beta}\:V_{(2)\beta}}}$ according to the notation used in papers \cite{A,ROMP} and equations (\ref{UO}-\ref{VO}).
The condition $[(1+C)^2-D^2]>0$ allows for two possible situations, $1+C > 0$ or $1+C < 0$. For the particular case when $1+C > 0$, the transformations (\ref{TN1}-\ref{TN2}) are manifesting that an electromagnetic gauge transformation on the vector field $A^{\alpha}$, that leaves invariant the electromagnetic field $f_{\mu\nu}$, generates a boost transformation on the normalized tetrad vector fields $\left({V_{(1)}^{\alpha} \over \sqrt{-V_{(1)}^{\beta}\:V_{(1)\beta}}}, {V_{(2)}^{\alpha} \over \sqrt{V_{(2)}^{\beta}\:V_{(2)\beta}}}\right)$. The case $1+C < 0$, represents the composition of two transformations. A full inversion of the normalized tetrad vector fields $\left({V_{(1)}^{\alpha} \over \sqrt{-V_{(1)}^{\beta}\:V_{(1)\beta}}},
{V_{(2)}^{\alpha} \over \sqrt{V_{(2)}^{\beta}\:V_{(2)\beta}}}\right)$, and a boost. If the case is that $[(1+C)^2-D^2]<0$, the vectors $V_{(1)}^{\alpha}$ and $V_{(2)}^{\alpha}$ will change their timelike or spacelike character,

\begin{eqnarray}
\tilde{V}_{(1)}^{\alpha}\:\tilde{V}_{(1)\alpha} &=&
[-(1+C)^2+D^2]\:(-V_{(1)}^{\alpha}\:V_{(1)\alpha})\label{FPI}\\
(-\tilde{V}_{(2)}^{\alpha}\:\tilde{V}_{(2)\alpha}) &=&
[-(1+C)^2+D^2]\:V_{(2)}^{\alpha}\:V_{(2)\alpha}\ .\label{SPI}
\end{eqnarray}

These are special improper transformations on blade one. They have the property of being a composition of boosts and a discrete transformation given by $\Lambda^{o}_{\:\:o} = 0$, $\Lambda^{o}_{\:\:1} = 1$, $\Lambda^{1}_{\:\:o} = 1$, $\Lambda^{1}_{\:\:1} = 0$. We notice that this discrete transformation is not a Lorentz transformation because it is a reflection. They might also be composed with a full inversion, see reference \cite{A,ROMP} for the whole analysis.  In the local blade or plane two, the choice $Y_{\alpha} = \ast A_{\alpha} + \ast \Lambda_{,\alpha}$ induces just local spatial rotation tetrad vector transformations on (\ref{V3}-\ref{V4}) or the normalized (\ref{ZO}-\ref{WO}).  There is a similar analysis for the vector transformations (\ref{ZO}-\ref{WO}) in the local plane two generated by ($Z^{\alpha}, W^{\alpha}$). In the local blade or plane two, the choice $Y_{\alpha} = \ast A_{\alpha} + \ast \Lambda_{,\alpha}$ induces just local spatial rotation tetrad vector transformations,

\begin{eqnarray}
{\tilde{V}_{(3)}^{\alpha}
\over \sqrt{\tilde{V}_{(3)}^{\beta}\:\tilde{V}_{(3)\beta}}}&=&
{(1+N) \over \sqrt{(1+N)^2+M^2}}
\:{V_{(3)}^{\alpha} \over \sqrt{V_{(3)}^{\beta}\:V_{(3)\beta}}} -
{M \over \sqrt{(1+N)^2+M^2}}
\:{V_{(4)}^{\alpha} \over \sqrt{V_{(4)}^{\beta}\:V_{(4)\beta}}}\label{TN3}\\
{\tilde{V}_{(4)}^{\alpha}
\over \sqrt{\tilde{V}_{(4)}^{\beta}\:\tilde{V}_{(4)\beta}}}&=&
{M \over \sqrt{(1+N)^2+M^2}}
\:{V_{(3)}^{\alpha} \over \sqrt{V_{(3)}^{\beta}\:V_{(3)\beta}}} +
{(1+N) \over \sqrt{(1+N)^2+M^2}}
\:{V_{(4)}^{\alpha} \over \sqrt{V_{(4)}^{\beta}\:V_{(4)\beta}}}\ ,
\label{TN4}
\end{eqnarray}

where,

\begin{eqnarray}
\tilde{V}_{(3)}^{\alpha}\:\tilde{V}_{(3)\alpha} &=&
[(1+N)^2+M^2]\:V_{(3)}^{\alpha}\:V_{(3)\alpha}\label{FPS}\\
\tilde{V}_{(4)}^{\alpha}\:\tilde{V}_{(4)\alpha} &=&
[(1+N)^2+M^2]\:V_{(4)}^{\alpha}\:V_{(4)\alpha}\ ,\label{SPS}
\end{eqnarray}

and where the relation $V_{(3)}^{\alpha}\:V_{(3)\alpha}=
V_{(4)}^{\alpha}\:V_{(4)\alpha}$ has been used with the following notation,

\begin{eqnarray}
M&=&(-Q/2)\:V_{(3)\sigma}\:\ast \Lambda^{\sigma} / (\:V_{(4)\beta}\:
V_{(4)}^{\beta}\:)\label{COEFFM}\\
N&=&(-Q/2)\:V_{(4)\sigma}\:\ast \Lambda^{\sigma} / (\:V_{(3)\beta}\:
V_{(3)}^{\beta}\:)\ .\label{COEFFN}
\end{eqnarray}

As long as $[(1+N)^2+M^2]>0$ the transformations (\ref{TN3}-\ref{TN4}) are manifesting that an electromagnetic gauge transformation on the vector field $\ast A^{\alpha}$ that leaves invariant the dual electromagnetic field $\ast f_{\mu\nu}$, generates a spatial rotation on the normalized tetrad vector fields $\left(Z^{\alpha}={V_{(3)}^{\alpha} \over \sqrt{V_{(3)}^{\beta}\:V_{(3)\beta}}}, W^{\alpha}={V_{(4)}^{\alpha} \over \sqrt{V_{(4)}^{\beta}\:V_{(4)\beta}}}\right)$. We reiterate that local tetrad electromagnetic gauge transformations can be interpreted as new or different gauge choices $X_{\alpha} = A_{\alpha} + \Lambda_{,\alpha}$ and $Y_{\alpha} = \ast A_{\alpha} + \ast \Lambda_{,\alpha}$.

Written in terms of these tetrad vectors, the electromagnetic field is,

\begin{equation}
f_{\alpha\beta} = -2\:\sqrt{-Q/2}\:\:\cos\alpha\:\:U_{[\alpha}\:V_{\beta]} +
2\:\sqrt{-Q/2}\:\:\sin\alpha\:\:Z_{[\alpha}\:W_{\beta]}\ .\label{EMF}
\end{equation}

Equation (\ref{EMF}) represents maximum simplification in the expression of the electromagnetic field. The true degrees of freedom are the local scalars $\sqrt{-Q/2}$ and $\alpha$. Local gauge invariance is manifested explicitly through the possibility of ``hyperbolic rotating'' through a scalar angle $\phi$ on blade one by a local gauge transformation (\ref{UT}-\ref{VT}) the tetrad vectors $U^{\alpha}$ and $V^{\alpha}$, such that $U_{[\alpha}\:V_{\beta]}$ remains invariant \cite{A}. Analogous for discrete transformations on blade one. Similar analysis on blade two. A ``spatial rotation'' of the tetrad vectors $Z^{\alpha}$ and $W^{\alpha}$ through an ``angle'' $\varphi$ as in (\ref{ZT}-\ref{WT}), such that $Z_{[\alpha}\:W_{\beta]}$ remains invariant \cite{A}. All this formalism clearly provides a technique to maximally simplify the expression for the electromagnetic field. The new expression for the metric tensor is,

\begin{equation}
g_{\alpha\beta} = -U_{\alpha}\:U_{\beta} + V_{\alpha}\:V_{\beta} +
Z_{\alpha}\:Z_{\beta} + W_{\alpha}\:W_{\beta}\ .\label{MT}
\end{equation}

The stress-energy tensor can be written,

\begin{equation}
T_{\alpha\beta} = (Q/2)\: \left[-U_{\alpha}\:U_{\beta} +
V_{\alpha}\:V_{\beta} -
Z_{\alpha}\:Z_{\beta} - W_{\alpha}\:W_{\beta}\right]\ .\label{SET}
\end{equation}

We proceed to apply all this geometrical elements to the Reissner-Nordstr\"{o}m case. The line element for this spacetime is given by the following expression \cite{RW,MC2,CBDW},

\begin{eqnarray}
ds^{2} = - (1 - {2m \over r} + {q^{2} \over r^{2}})\: dt^{2} + (1 - {2m \over r} + {q^{2} \over r^{2}})^{-1}\: dr^{2} + r^{2}\:(d\theta^{2} + \sin^{2}\theta\:d\phi^{2})\ . \label{reissnord}
\end{eqnarray}

We choose $X^{\rho} = A^{\rho}$ and $Y^{\rho} = \ast A^{\rho}$, where the symbol $\ast$ in this particular last case is not the Hodge operator but a name. In the standard spherical coordinates $t, r, \theta, \phi$ the only non-zero components for the potentials will be $A_{t} = - q / r$ and $\ast A_{\phi} = -q\:\cos\theta$. With these potential components we find that the only non-zero components for the electromagnetic tensor $f_{\mu\nu} = A_{\nu ;\mu} - A_{\mu ;\nu}$ and its Hodge dual $\ast f_{\mu\nu} = \ast A_{\nu ;\mu} - \ast A_{\mu ;\nu}$ are $f_{tr} = - q / r^{2}$ and $\ast f_{\theta\phi} = q\:\sin\theta$. The symbol $;$ stands for covariant derivative with respect to the metric tensor $g_{\mu\nu}$, which in our case is the Reissner-Nordstr\"{o}m geometry. It is easy to check that the only non-zero components of the extremal field and its dual are $\xi_{tr} = f_{tr}$ and $\ast\xi_{\theta\phi} = \ast f_{\theta\phi}$. The reason for this is due to the fact that for the Reissner-Nordstr\"{o}m geometry $\tan(2\alpha)=0$. The extremal field and its dual are gauge invariants, therefore their expression is unique. When we observe the skeletons in tetrad vectors (\ref{U}-\ref{W}), we notice that these skeletons are unique. Precisely because of their local gauge invariance, and also because they also diagonalize locally and covariantly  the stress-energy tensor in a unique fashion. Then, it becomes evident that we need both the Coulomb and the Monopole field in order to implement their construction. Both electromagnetic fields simultaneously. We proceed again to write explicitly the only non-zero components of vectors (\ref{U}-\ref{W}),

\begin{eqnarray}
U^{t} &=& - (\sqrt{q^{2}}/q) / \sqrt{1 - {2m \over r} + {q^{2} \over r^{2}}} \label{Ut}\\
V^{r} &=& \sqrt{1 - {2m \over r} + {q^{2} \over r^{2}}} \label{Vr}\\
Z^{\theta} &=& - \sqrt{\cos^{2}\theta} / (r\:\cos\theta)  \label{Ztheta}\\
W^{\phi} &=& - \sqrt{q^{2}}\:\sqrt{\cos^{2}\theta} / (q\:r\:\sin\theta\:\cos\theta) \ .\label{Wphi}
\end{eqnarray}

In this particular coordinate system we need to be careful because both vectors $V_{(3)}^{\alpha}$ and $V_{(4)}^{\alpha}$ before normalizing would be zero at the coordinate value $\theta = \pi / 2$. Since the purpose of this section is not to find suitable coordinate coverings but to show that both the Coulomb and the Monopole electromagnetic fields are indispensable components of the tetrad vectors that make up the Reissner-Nordstr\"{o}m geometry, we will not search for other coordinate coverings. Please see reference \cite{AMONO} for a detailed tetrad analysis of the Reissner-Nordstr\"{o}m case.

In section \ref{equivalence} we will prove the equivalence between the local inertial frames and local gauge states of the electromagnetic field for the tetrad that locally and covariantly diagonalizes the stress-energy tensor. In section \ref{equiLorentz} we will generalize the proof to any locally Lorentz transformed tetrad. For the particular case when the electromagnetic field is null, please see section \ref{sec:appI}.

\section{Equivalence for the tetrad that diagonalizes the stress-energy tensor}
\label{equivalence}

The theorem proved in manuscript \cite{A} for blade one states that there is an isomorphism between the local electromagnetic gauge group of transformations and the local group LB1, essentially the local boosts on blade one and two kinds of discrete transformations, see reference \cite{A}. Therefore, to each local gauge state of the electromagnetic field corresponds either a local boost of the two local tetrad vectors that span plane one, that is vectors (\ref{UO}-\ref{VO}), or a discrete transformation of them. These all means that locally, to each absolute value of a velocity corresponds a unique electromagnetic gauge. For local Lorentz boosts on the plane one. Let us suppose then that an object that might be a classical object is boosted by a mechanical means. For example if the object is a starship floating in outer space far away from any star or planet when engine jets are started and the velocity increases from zero to some value $v$ and then the engines are at that point shut off from the point of view of some observer. The starship will continue moving at a constant velocity $v$ for this observer. Every atom in the starship structure is made of electrons and nuclei. The nuclei are made of protons and neutrons. Electrons have an associated non-null electromagnetic field as well as the proton. Even the quarks making up the proton and the neutron have associated non-null electromagnetic fields. From equations (\ref{TN1}-\ref{TN2}) we can see that $\cosh\zeta=\frac{1}{\sqrt{1-\tanh^{2}\zeta}}={(1+C) \over \sqrt{(1+C)^2-D^2}}$ where $\beta=\tanh\zeta$ and $\beta=\frac{v}{c}$. The variable $\zeta$ is the rapidity. In the boost case $1+C>0$ where the observer measures $\beta>0$ then we can write $\beta=\tanh\zeta=|D|/(1+C)$. The key to the result that follows is that there is an isomorphism between the local electromagnetic gauge group $U(1)$ and the local group of tetrad transformations LB1. LB1 is composed by $SO(1,1)$ and two discrete transformations as explained above. Therefore, there must be a local electromagnetic gauge transformation that is mapped into the boost that corresponds to the velocity $\beta$. There is a local $\partial_{\mu}\Lambda=\Lambda_{\mu}$ corresponding to $\beta$. We conclude that boosting the matter in the starship to velocity $\beta$ is indistinguishable from changing the local electromagnetic gauge inside the matter structure by the local gradient $\partial_{\mu}\Lambda=\Lambda_{\mu}$ according to previous established notation in section \ref{intro}. We must emphasize that the microparticles components of matter might also have associated to them $SU(2)$, $SU(3)$ and $SU(N)$ local gauge fields. Analogously to the electromagnetic case described so far it is possible to develop a whole theory of local tetrads associated to these Yang-Mills fields, see references \cite{AYM,gaugeinvmeth,A3,ASU3,ASUN,Sl2C}. It is proven that the local gauge groups for Yang-Mills non-Abelian theories are isomorphic to tensor products of LB1 groups or independently to tensor products of LB2 groups of local tetrad transformations. By following a completely similar reasoning to the electromagnetic case we would conclude that boosting the matter in the starship to velocity $\beta$ is indistinguishable from changing the local Yang-Mills gauge fields inside the matter structure by the local $S$ and the local gradient $\partial_{\mu}S$ where $S$ is generic notation for a local non-Abelian gauge transformations in any of the local groups $SU(2) \times U(1)$,  $SU(3) \times SU(2) \times U(1)$ or even $SU(N) \times SU(N-1) \times \cdots \times SU(3) \times SU(2) \times U(1)$, see references \cite{AYM,gaugeinvmeth,A3,ASU3,ASUN,Sl2C,LE}.

\section{Equivalence for Lorentz transformed tetrads}
\label{equiLorentz}

However, the point remains to be proved that there is a similar relationship for a locally Lorentz transformed tetrad, such that the new plane or blade one will be Lorentz transformed with respect to the one that diagonalizes the stress-energy tensor. We proceed then to call generically the tetrad set (\ref{U}-\ref{W}) by the standard name $E_{\alpha}^{\:\:\mu}$. For the second electromagnetic tetrad we will need a local Lorentz transformation. Let us analyze the expression $\tilde{E}_{\delta}^{\:\:\rho} = \Lambda^{\alpha}_{\:\:\delta}\:E_{\alpha}^{\:\:\rho}$. This will be a Lorentz transformed electromagnetic tetrad vector. Then, keeping the same notation as in section VI of reference \cite{A}, we call,

\begin{eqnarray}
\tilde{\xi}^{\mu\nu} &=& -2\:\sqrt{-Q/2}\:\Lambda^{\delta}_{\:\:o}\:\Lambda^{\gamma}_{\:\:1}\:E_{[\delta}^{\:\:\mu}\:E_{\gamma]}^{\:\:\nu}\label{ET}\\
\ast \tilde{\xi}^{\mu\nu} &=& 2\:\sqrt{-Q/2}\:\Lambda^{\delta}_{\:\:2}\:\Lambda^{\gamma}_{\:\:3}\:E_{[\delta}^{\:\:\mu}\:E_{\gamma]}^{\:\:\nu}\ .\label{DET}
\end{eqnarray}

Now, with these fields, the $\tilde{\xi}_{\mu\nu}$, and its dual $\ast \tilde{\xi}_{\mu\nu}$, we can repeat the procedure followed in \cite{A}, and the transformed tetrads $\tilde{E}_{\alpha}^{\:\:\rho}$, can be rewritten completely in terms of these ``new'' extremal fields. It is straightforward to prove that $\tilde{\xi}^{\mu\lambda}\:\ast \tilde{\xi}_{\mu\nu} = 0$. It is also evident that $\tilde{E}_{o}^{\:\:\mu} \:\ast \tilde{\xi}_{\mu\nu} = 0 = \tilde{E}_{1}^{\:\:\mu} \:\ast \tilde{\xi}_{\mu\nu}$. Therefore $\tilde{E}_{o}^{\:\:\mu}$ and $\tilde{E}_{1}^{\:\:\mu}$ belong to the plane generated by the normalized version of vectors like $\tilde{\xi}^{\mu\nu}\:\tilde{\xi}_{\lambda\nu}\:\tilde{X}^{\lambda}$ and $\tilde{\xi}^{\mu\nu}\:\tilde{X}_{\nu}$.
Then, for instance we will be able to write the timelike $\tilde{E}_{o}^{\:\:\mu}$ as the the normalized version of the timelike $\tilde{\xi}^{\mu\nu}\:\tilde{\xi}_{\lambda\nu}\:\tilde{X}^{\lambda}$ for some vector field  $\tilde{X}^{\lambda}$. We remind ourselves that the relation between the normalized versions of the two vectors that locally determine blade one, $\tilde{\xi}^{\mu\nu}\:\tilde{\xi}_{\lambda\nu}\:X^{\lambda}$ and  $\tilde{\xi}^{\mu\nu}\:X_{\nu}$ on one hand, and $\tilde{\xi}^{\mu\nu}\:\tilde{\xi}_{\lambda\nu}\:\tilde{X}^{\lambda}$ on the other hand, is established through a LB1 gauge transformation \cite{A} on the vector $X^{\lambda}\longrightarrow \tilde{X}^{\lambda}=X^{\lambda}+ \Lambda^{,\lambda}$. Analogous analysis for $\tilde{E}_{2}^{\:\:\mu}$ and $\tilde{E}_{3}^{\:\:\mu}$ on blade two.  Gauge transformations of the electromagnetic tetrads we remind ourselves are nothing but a special kind of tetrad transformations that belong either to the groups LB1 or LB2. This method essentially says that the local Lorentz transformation of the electromagnetic tetrads is structure invariant, or construction invariant. This means that after a Lorentz transformation we can manage to rewrite the new transformed tetrads using skeletons and gauge vectors following the same pattern as for the original tetrad before the Lorentz transformation. We will call this property tetrad structure covariance. Therefore, we next proceed to write the four orthonormal vectors $\tilde{E}_{\delta}^{\:\:\rho}$,

\begin{eqnarray}
\tilde{U}^{\alpha} &=& \tilde{\xi}^{\alpha\lambda}\:\tilde{\xi}_{\rho\lambda}\:\tilde{X}^{\rho} \:
/ \: (\: \sqrt{-\tilde{Q}/2} \: \sqrt{\tilde{X}_{\mu} \ \tilde{\xi}^{\mu\sigma} \
\tilde{\xi}_{\nu\sigma} \ \tilde{X}^{\nu}}\:) \label{ULT}\\
\tilde{V}^{\alpha} &=& \tilde{\xi}^{\alpha\lambda}\:\tilde{X}_{\lambda} \:
/ \: (\:\sqrt{\tilde{X}_{\mu} \ \tilde{\xi}^{\mu\sigma} \
\tilde{\xi}_{\nu\sigma} \ \tilde{X}^{\nu}}\:) \label{VLT}\\
\tilde{Z}^{\alpha} &=& \ast \tilde{\xi}^{\alpha\lambda} \: \ast \tilde{Y}_{\lambda} \:
/ \: (\:\sqrt{\ast \tilde{Y}_{\mu}  \ast \tilde{\xi}^{\mu\sigma}
\ast \tilde{\xi}_{\nu\sigma}  \ast \tilde{Y}^{\nu}}\:)
\label{ZLT}\\
\tilde{W}^{\alpha} &=& \ast \tilde{\xi}^{\alpha\lambda}\: \ast \tilde{\xi}_{\rho\lambda}
\:\ast \tilde{Y}^{\rho} \: / \: (\:\sqrt{-\tilde{Q}/2} \: \sqrt{\ast \tilde{Y}_{\mu}
\ast \tilde{\xi}^{\mu\sigma} \ast \tilde{\xi}_{\nu\sigma} \ast \tilde{Y}^{\nu}}\:) \ .
\label{WLT}
\end{eqnarray}

In order to prove the properties of the tetrad set (\ref{ULT}-\ref{WLT}) it is just necessary to transcribe many of the results introduced in section \ref{intro}. We are assuming that our choice for vectors $\tilde{X}^{\rho}$ and $\tilde{Y}^{\rho}$ are not making the tetrad trivial. Now, and this is the point of this section, if we choose $\tilde{X}^{\rho} = A^{\rho}$ and $\tilde{Y}^{\rho} = \ast A^{\rho}$ and introduce local transformations $A_{\alpha} \rightarrow A_{\alpha} + \Lambda_{,\alpha}$ and $\ast A_{\alpha} \rightarrow \ast A_{\alpha} + \ast \Lambda_{,\alpha}$ such that the new extremal fields $\tilde{\xi}_{\mu\nu}$ and its dual $\ast \tilde{\xi}_{\mu\nu}$ remain invariant, then the results of section \ref{equivalence} are reproduced once again. One might ask about the local choice of vectors $X^{\mu}$ and $Y^{\mu}$ hidden in the tetrads $E_{\alpha}^{\:\:\mu}$ or equivalently the tetrad vectors (\ref{U}-\ref{W}). Because these tetrad vectors are hidden in $\tilde{\xi}_{\mu\nu}$ and its dual $\ast \tilde{\xi}_{\mu\nu}$. How we manage to transform the gauge vectors $A_{\alpha} \rightarrow A_{\alpha} + \Lambda_{,\alpha}$ and $\ast A_{\alpha} \rightarrow \ast A_{\alpha} + \ast \Lambda_{,\alpha}$ without affecting the new extremal fields  $\tilde{\xi}_{\mu\nu}$, and its dual $\ast \tilde{\xi}_{\mu\nu}$, that we claim will remain invariant. One simple local gauge choice for them would be for instance $X^{\rho} = Y^{\rho} = \alpha_{,\nu}\:g^{\nu\rho}$, where $\alpha$ is the local complexion scalar defined in section \ref{intro}. It is a local gauge invariant, and this choice solves the problem with the local invariance of the new extremal fields. In fact, any local gauge invariant scalar would do the job like $Q_{,\nu}\:g^{\nu\rho}$, for instance. Returning to our issue of the equivalence of the local inertial frames and gauge on a new plane one, which is the result of a local Lorentz transformation of the plane one that ``diagonalizes'' the stress-energy tensor, the same conclusions that were reached for the plane one that ``diagonalizes'' the stress-energy tensor, are reached for the new plane. Locally, to each absolute value of a velocity there corresponds a unique electromagnetic gauge. Since the local Lorentz transformation is generic, we conclude that locally, to each absolute value of any velocity (less than c, of course) in any direction there corresponds an electromagnetic gauge. If we pick any local plane one, the relationship between velocity absolute value and electromagnetic gauge is one to one.

\section{Conclusions}
\label{conclusions}

We will begin the summary by putting together all the previous results in a theorem,

\newtheorem {guesslb1} {Theorem}
\begin{guesslb1}
Local inertial frames and electromagnetic gauge in Einstein-Maxwell theories are equivalent in the sense of isomorphic. \end{guesslb1}

As analyzed in reference \cite{ROMP} it is a piecewise isomorphism. Because we have in the local plane one boosts which are hyperbolic
rotations, boosts composed with full inversions which are the composition of two reflections, boosts composed with spacetime reflections and boosts composed with full inversions and spacetime reflections. In the local plane two we have spatial rotations. It is in a general sense a piecewise isomorphism. It is not only interesting but also surprising that the local inertial frames \cite{ICW} are even related locally to the electromagnetic gauge. Not only that, there are on each local orthogonal plane, a one to one correspondence to the special groups of tetrad transformations LB1 and LB2. This result is not trivial. It goes to the heart of a unified structure involving local inertial frames and gauge. The new tetrads introduced in manuscript \cite{A} make this relationship to become evident. Several properties of these tetrads are remarkable. For instance their skeleton-gauge vector structure. Their structure covariant or structure invariant nature under local Lorentz transformations as in section \ref{equiLorentz}. The fact that they allow to prove that the local electromagnetic gauge group is both isomorphic to the local groups LB1 and LB2, see references \cite{A,ROMP,SING}. This tetrad introduces maximal simplification in the expression of the electromagnetic field. Automatically diagonalizes locally and covariantly the stress-energy tensor. It is truly outstanding. What is even more outstanding is that we have proved either for Abelian or Yang-Mills non-Abelian local gauge groups of transformations that inertia or just a classical object boosted at velocity $\beta$ is indistinguishable from changing the local gauge inside the matter structure by local gauge transformations, Abelian and non-Abelian. At the core of all these demonstrations lies the fact that the no-go theorems \cite{CMNG,SWNG,LORNG}  from the sixties are incorrect. We read from reference \cite{CMNG} ``S (the scattering matrix) is said to be Lorentz-invariant if it possesses a symmetry group locally isomorphic to the Poincar\`{e} group P.\ldots A symmetry transformation is said to be an internal symmetry transformation if it commutes with P. This implies that it acts only on particle-type indices, and has no matrix elements between particles of different four-momentum or different spin. A group composed of such transformations is called an internal symmetry group''. The local electromagnetic gauge group of transformations $U(1)$ has been proven to be isomorphic to local groups of tetrad transformations LB1 and LB2 on both the local orthogonal planes of Einstein-Maxwell stress-energy eigenvectors. The local group of electromagnetic gauge transformations is isomorphic on the local plane one to the local group LB1. The group LB1 is given by $SO(1,1) \times Z_{2} \times Z_{2}$ where $SO(1,1)$ is proper orthochronous. The first $Z_{2}$ is given by $\{I_{2 \times 2}, -I_{2 \times 2}\}$ and the second $Z_{2}$ is given by $\{I_{2 \times 2}, \mbox{the swap}\: (01|10)\}$. There are two discrete transformations. One of them that we designated as $-I_{2 \times 2}$ is the full inversion two by two and the other a reflection designated as $\mbox{the swap}\: (01|10)$ and given by $\Lambda^{o}_{\:\:o} = 0$, $\Lambda^{o}_{\:\:1} = 1$, $\Lambda^{1}_{\:\:o} = 1$,  $\Lambda^{1}_{\:\:1} = 0$ which is not a Lorentz transformation. We would have to add in order to complete the image of the map $SO(1,1) \times Z_{2} \times Z_{2}\: \bigoplus \: \{light\:cone\:gauge\}$ where the light cone gauge includes the inhomogeneous two solutions to the differential equations in the local future and past light cones established in reference \cite{SING} where the reflection through the asymptote $Y=X$ will produce two more identical inhomogeneous solutions. A total of four. When we talk about the light cone gauge we are not referring to this concept in the sense of book \cite{MK} or manuscript \cite{JSJHS}. Light cone gauge means that when we map the tetrad vectors in the local plane one through specific local electromagnetic gauge transformations the vectors that result of these transformations are null vectors. In manuscript \cite{SING} the general case has been presented as well as examples in the Coulomb and Reissner Nordstr\"{o}m geometries. On the local orthogonal plane two and independently from the mapping in the plane one, the local group of electromagnetic gauge transformations is mapped onto the local group of spatial tetrad rotations $SO(2)$. For example in reference \cite{CMNG} we can read ``Let $G$ be a connected symmetry group of the $S$ matrix, and let the following five conditions hold: 1. (Lorentz invariance) $G$ contains a subgroup locally isomorphic to $P$ (Poincar\'{e} group). 2. (Particle-finiteness) \ldots ''. If the local gauge groups of the standard model are isomorphic to the local tetrad groups of transformations LB1 or LB2 in the Abelian electromagnetic case \cite{A,ROMP,SING,ATGU,GLAW} or tensor products of them in the Yang-Mills general case \cite{AYM,gaugeinvmeth,A3,ASU3,ASUN} then the subgroup is $G$ itself and the no-go theorems are void of any content and incorrect. The tetrad vectors in both mappings never leave their local original planes that they span thus becoming manifest symmetries of the metric tensor. Both internal-spacetime mappings have been proved to be isomorphisms, see references \cite{A,ROMP,SING}. These local groups of transformations LB1 and LB2$=SO(2)$ are composed of Lorentz transformations and even though the LB1 improper discrete reflection flip is not a Lorentz transformation, it is composed with this exception of spacetime Lorentz transformations, see reference \cite{A}. The spacetime flip is a discrete transformation given by $\Lambda^{o}_{\:\:o} = 0$, $\Lambda^{o}_{\:\:1} = 1$, $\Lambda^{1}_{\:\:o} = 1$, $\Lambda^{1}_{\:\:1} = 0$. We notice that this discrete transformation is not a Lorentz transformation because it is a reflection. Therefore the local Lorentz group of spacetime transformations cannot commute with LB1 or LB2 since Lorentz transformations on a local plane do not commute with Lorentz transformations on another local different plane necessarily. Therefore, since the local internal groups of transformations do not necessarily commute with the local Lorentz transformations because they are isomorphic to local groups of tetrad spacetime transformations on local orthogonal special planes, which are unique, the no-go theorems are incorrect. Similar results were proven for the Yang-Mills cases $SU(2) \times U(1)$,  $SU(3) \times SU(2) \times U(1)$ or even $SU(N) \times SU(N-1) \times \cdots \times SU(3) \times SU(2) \times U(1)$ in geometrodynamics, see references \cite{AYM,gaugeinvmeth,A3,ASU3,ASUN,Sl2C,LE}. In addition we have proven that all of these results are still valid when spacetime is under perturbations but locally and instantaneously. The perturbative classical Abelian case has already been discussed in manuscript \cite{dsmg,PIRT2019} and the Yang-Mills perturbative  case in reference \cite{DSBYM}. The local planes of diagonalization when a system is under perturbations, tilt. The local planes of gauge symmetry which are the local planes of stress-energy eigenvectors will tilt with the perturbations and the symmetries will become instantaneous. The results in section \ref{equivalence} will be valid but instantaneously. We quote from \cite{Weyl} ``Here is not the place to write down the Lorentz transformations and to sketch how special relativity theory with its fixed causal and inertial structure gave way to general relativity where these structures have become flexible by their interaction with matter. I only want to point out that it is the inherent symmetry of the four-dimensional continuum of space and time that relativity deals with. We found that objectivity means invariance with respect to the group of automorphisms''. We also quote H. Weyl from \cite{KC} ``By this new situation, which introduces an atomic radius into the field equations themselves -but not until this step- my principle of gauge-invariance, with which I had hoped to relate gravitation and electricity, is robbed of its support. But it is now very agreeable to see that this principle has an equivalent in the quantum-theoretical field equations which is exactly like it in formal respects; the laws are invariant under simultaneous replacement of $\psi$ by $\exp(\imath h\lambda)\:\psi$, $\phi_{\alpha}$ by $\phi_{\alpha} - {\partial\lambda \over \partial x^{\alpha}}$, where $\lambda$ is an arbitrary real function of position and time. Also the relation of this property of invariance to the law of conservation of electricity remains exactly as before \ldots the law of conservation of electricity ${\partial\rho^{\alpha} \over \partial x^{\alpha}} = 0$ follows from the material as well as from the electromagnetic equations. The principle of gauge-invariance has the character of general relativity since it contains an arbitrary function $\lambda$, and can certainly be understood in terms of it''.

\section{Appendix I}
\label{sec:appI}

In this section we will discuss the null electromagnetic field. The contracted curvature tensor will be null even though the curvature tensor itself will not be null,

\begin{eqnarray}
R_{\mu\nu}\:R^{\mu\nu}=0 \ .\label{NULLCURVATURE}
\end{eqnarray}

After solving these equations in a Minkowski reference system with signature $(-,+,+,+)$, please see reference \cite{MW} pages 545-547 for the detailed proof, we arrive at the conclusion that $R_{\mu\nu}=2\:k_{\mu}\:k_{\nu}$ where $k_{\mu}=(-\kappa,\kappa,0,0)$ and $k_{\mu}\:k^{\mu}=0$. This analysis is covariant therefore valid in any reference system. As stated in reference \cite{MW} there is no Lorentz transformation that will diagonalize a null Ricci tensor or any Lorentz transformation that will parallelize electric and magnetic fields $({\bf e},{\bf h})$ that satisfy the null condition ${\bf e}. {\bf h}=0={\bf h}^{2}-{\bf e}^{2}$. After several disquisitions in reference \cite{MW} it is found that the Maxwell root of equations $R_{\mu\nu}=2\:k_{\mu}\:k_{\nu}$ where $k_{\mu}=(-\kappa,\kappa,0,0)$ will be provided by,

\begin{eqnarray}
f_{\mu\nu}= k_{\mu}\:v_{\nu} - k_{\nu}\:v_{\mu}\ .\label{nullemfield}
\end{eqnarray}

where $v_{\nu}=(0,0,1,0)$, $v_{\nu}\:v^{\nu}=1$ and $k_{\mu}\:v^{\mu}=0$. The tensor (\ref{nullemfield}) is a Maxwell square root of the null tensor $R_{\mu\nu}$ and apart from a duality rotation is the only square root of $R_{\mu\nu}$. The reduced electromagnetic field (\ref{nullemfield}) is a null field in the Minkowski reference frame which has been used so far. Therefore it is also a null field in any other reference frame since the analysis has been covariant. The result of a duality rotation on the tensor (\ref{nullemfield}) can be presented as follows. $k_{\mu}$ will be left unchanged while the vector $v_{\nu}$ is rotated about $k_{\mu}$. Nonetheless the vector $v_{\nu}$ is not determined by the tensor (\ref{nullemfield}) since a transformation $v_{\nu} \rightarrow v_{\nu} + a\:k_{\nu}$ with $a$ a constant, gives the same field tensor (\ref{nullemfield}) and satisfies also the equations $v_{\nu}\:v^{\nu}=1$ and $k_{\mu}\:v^{\mu}=0$.

\section{Conflict of interests statement}
\label{interest}

The authors declare that they have no known competing financial interests or personal relationships that could have appeared to influence the work reported in this paper.

\section{Data and materials availability statement}
\label{data}

There is no data to be reported in this paper.

\section{Ethical Approval}
\label{interest}

This declaration is ``not applicable''.

\section{Funding}
\label{interest}

There is no funding to report.




\end{document}